\documentclass[12pt,aps,pre]{revtex4-1}
\usepackage{graphicx}
\usepackage{graphics}
\usepackage{indentfirst}
\usepackage{subfigure}
\usepackage{amsmath, amsthm, amssymb}
\usepackage[T1]{fontenc}
\usepackage[latin9]{inputenc}

\begin{document}

\title{Phase space flow in the Husimi representation}

\author{M.~ Veronez and M.~A.~M.~de Aguiar}

\affiliation{Instituto de F\'{\i}sica `Gleb Wataghin', Universidade
Estadual de Campinas (UNICAMP)\\ 13083-970, Campinas, Brazil}

\begin{abstract}

We derive a continuity equation for the Husimi function evolving under a general
non-hermitian Hamiltonian and identify the phase space flow associated with it.
For the case of unitary evolution we obtain explicit formulas for the quantum
flow, which can be written as a classical part plus semiclassical corrections. 
These equations are the analogue of the Wigner flow, which displays several 
non-intuitive features like momentum inversion and motion of stagnation points. 
Many of these features also appear in the Husimi flow and, therefore, are not related 
to the negativity of the Wigner function as previously suggested. We test the exact 
and semiclassical formulas for a particle in a double well potential. We find that the 
zeros of the Husimi function are saddle points of the flow, and are always followed 
by a center. Merging or splitting of stagnation points, observed in the Wigner flow,
does not occur because of the isolation of the Husimi zeros.

\end{abstract}

\maketitle

\section{Introduction}

In classical mechanics the state of a system is often associated to a point in phase
space. The initial condition defined by this point specifies a unique trajectory that
guides the evolution of the system. This association, however, is not accurate
in many situations due to imprecisions in assessing the system's state or the
statistical nature of the problem at hand. In these cases it is better to work
with probability distributions and the associated Liouville equation than with
individual trajectories and Hamilton's equations. 

In one-dimension the phase space is constructed with a pair of canonically
conjugate variables, position $x$ and momentum $p$, and the classical dynamics of a
function $F\left(x,p;t\right)$ can be written in the form of a continuity
equation
\begin{equation}
\frac{\partial F}{\partial t}+\mathbf{\nabla\cdot J}_{cl}=0
\label{eq:continuity_equation}
\end{equation}
where $\mathbf{\nabla} = \left(\frac{\partial}{\partial x},
\frac{\partial}{\partial p}\right)$, the classical current vector is given by
\begin{equation}
\mathbf{J}_{cl}=\left(\begin{array}{c}
J_{x}\\
J_{p}
\end{array}\right)=\left(\begin{array}{c}
\dot{x}F\\
\dot{p}F
\end{array}\right),\label{eq:classical_currents_1}
\end{equation}
and the dots indicate total derivative with respect to time. The characteristic
property of this construction is that each point of the phase space on which $F$
is evaluated is guided by a well-defined trajectory and the flow lines of the
current are just the tangent vectors to these trajectories. The dynamics of the
function $F$ is thus trivial in the sense that the points just follow the flow,
that is, $F\left(x,p;t\right)=F\left(x_0,p_0;0\right)$ where $x_0$ and $p_0$ are
the initial conditions that propagate to $x$ and $p$ in the time $t$.

In quantum mechanics states are naturally described in terms of probability
distributions but, due to the uncertainty principle, phase space
representations have interpretations that are different from their classical
counterparts. Two of the most used quantum phase space representations are the
Wigner and the Husimi functions. The Wigner function $W(x,p,t)$ associated to a
pure state $|\psi \rangle$ has the correct marginal probability distributions
when projected into the $x$ or $p$ subspaces, but can be itself negative. The
Husimi distribution $Q\left(x,p,t\right)$, on the other hand, is positive by
definition, but does not project onto the correct marginal distributions.
Despite these well known properties, both representations, and others as well,
have been successful employed in many quantum mechanical treatments
\cite{glauber,cohen,hillery,kla,perel}, particularly in the study of the
boundary between quantum and classical mechanics
\cite{hell75,hk84,berry89,kurchan89,xavier97,kay94a,kay94b,ozorio98,bar01,
mill01,pollak03,rib04,sha04,sha08,agui10,thiago11a,thiago11b}.

If $F(x,p;t)$ is a phase space representation of a quantum state, a natural
question to ask is whether it obeys a continuity equation similar to
(\ref{eq:continuity_equation}) and if a flow can be defined. The difficulty
resides in the uncertainty principle, that forbids the definition of authentic
quantum trajectories $x\left(t\right)$ and $p\left(t\right)$ guiding the
dynamics in these representations. Although it may seem conflicting, it has been
previously shown that the dynamics of the Wigner function can indeed be
cast as a continuity-like equation \cite{donoso03,bauke11,skr13}, thus
confirming that for this particular representation a flow \emph{is} well-defined
even though the trajectories in the classical sense \emph{are} \emph{not}.

The quantum flow associated with the Wigner function exhibits many interesting
non-classical features, like travelling stagnation points that can merge with or
split from other such points, vortices and conservation of the flow winding
number \cite{skr13}. It was argued that many of these complex features were
consequences of the negativity of the Wigner function. 

In this work we employ the coherent state representation and develop a flow
formalism for the Husimi function. A continuity equation for the Husimi dynamics
has already been demonstrated but only for a particular class of systems
\cite{skodje89}. Here we derive formulas for the Husimi flow for very general
Hamiltonian systems and compare features of this flow with those of the Wigner
function.

We show that for Hermitian Hamiltonians there are no source or sink terms, which
do appear for non-unitary evolution. Like its Wigner counterpart, non-local
features lead to noticeable time-dependent distortions with respect to the
classical flow lines, including the displacement and motion of the classical
stability points and inversion of momentum. In \cite{skr13} it was reasoned that
inversion of momentum lines of the flow for the Wigner function was caused by
negativity of the function, which is itself a mark of non-classicality. The
inversion of momentum lines was also found here, for a positive definite
function, implying that such inversions are a more robust sign of quantumness
than negativity for some particular representation. For the considered example
we also found that every zero of the Husimi function behaves as a saddle point
of the currents, with no other saddles identified beside these zeros. 

The paper is organized as follows: in section II we define the Husimi function
and construct the associated continuity equation for general non-Hermitian 
Hamiltonians. We provide a detailed derivation of the flow and then restrict the
calculations to the unitary case. In section III we show how to obtain the
classical equation of motion given in terms of Poisson brackets from the quantum
flow and also derive semiclassical corrections to the classical flow. In section
IV we illustrate these features using as example a one-dimensional double well
potential. In section V we make some final remarks about the main results.

\section{The Husimi flow}

\subsection{The Husimi function}

The coherent states of a harmonic oscillator with mass $m$ and frequency
$\Omega$ are defined as the eigenstates of the annihilation operator
$\widehat{a}=\left(\gamma\widehat{x}+i\widehat{p}/\gamma\right)/\sqrt{2\hbar}$,
where $\gamma=\sqrt{m\Omega}$:
\begin{equation}
\widehat{a}\vert z\rangle=z\vert z\rangle.
\label{eq:coherent_state_eigen}
\end{equation}
The normalized coherent states can also be written as
\begin{equation}
\vert z\rangle=e^{-\frac{1}{2}\left|z\right|^{2}}e^{z\widehat{a}^{\dagger}}\vert0\rangle,
\label{eq:coherent_state_vacuum}
\end{equation}
which will be useful in what follows. Here $\vert 0\rangle$ is the ground state
of the oscillator, which is also the coherent state labeled by $z=0$. The
eigenvalue $z$ is generally complex and defines a one-dimensional complex
manifold $\Xi_{z}$ with volume element 
$\text{d}^{2}z\equiv\text{d}z^{\star}\wedge\text{d}z/2\pi i$ \cite{kla}. These
states form a basis and 
\begin{equation}
\widehat{1}=\int\text{d}^{2}z\vert z\rangle\langle z\vert.
\label{eq:identity_z}
\end{equation}

The coherent states can also be described as the ground state displaced by a
complex amount $z$ in the eigenvalue manifold $\Xi_{z}$. The ground state $\vert
0\rangle$ is a minimum uncertainty gaussian wavepacket in both momentum and
position representation, and so are the displaced states $\vert z\rangle$
\cite{perel}. The relation between $z$ and the mean momentum $p$ and position
$x$ of the displaced state is given by
\begin{equation}
z=\frac{1}{\sqrt{2\hbar}}\left(\gamma x+\frac{ip}{\gamma}\right).
\label{eq:z_position_momentum}
\end{equation}
This expression provides a bijection from the classical phase space $\Xi_{cl}$
of the variables $x$ and $p$ to the complex manifold $\Xi_{z}$
which parametrize the set of states $\vert z\rangle$.

Given a pure one-particle quantum state $\vert\psi\rangle$ and its projection 
$\psi\left(z^{\star},z\right)=\langle z\vert\psi\rangle$, the Husimi
function is the quasi-probability density associated to this wavefunction
under the volume form $\text{d}^{2}z$:
\begin{equation}
Q_{\psi}\left(z^{\star},z\right)=\left|\psi\left(z^{\star},z\right)\right|^{2} =
\text{tr}\left(\vert z\rangle\langle z\vert\widehat{\rho}\right)
\label{eq:Husimi_square_modulus}
\end{equation}
where $\widehat{\rho}=\vert\psi\rangle\langle\psi\vert$ is the density matrix.
This function is also called the $Q$-representation or $Q$-symbol of the quantum
state. The Husimi function is a positive function over the manifold $\Xi_z$, but
it is not a probability distribution, since the coordinate and momentum
probability densities cannot be retrieved as the marginals distributions from
Eq. (\ref{eq:Husimi_square_modulus}).

\subsection{Time dependent states}

Consider a time dependent quantum state $\vert\psi;t\rangle$ evolved under the
action of a time independent hamiltonian $\widehat{H}$ (which we do not assume
hermitian at this point),
\[
\vert\psi;t\rangle=\widehat{K}\left(t-t_{0}\right)\vert\psi;t_{0}\rangle,
\]
where $\widehat{K}=e^{ -i\widehat{H}t/\hbar}$ is the time evolution operator.
The Husimi function inherits the time dependence of the state
\begin{equation}
Q_{\psi}\left(z^{\star},z;t\right)=\text{tr} 
\left(\vert z\rangle\langle z\vert\widehat{\rho}\left(t\right)\right),
\label{eq:Husimi_trace_TD}
\end{equation}
where $\widehat{\rho}\left(t\right)=\vert\psi;t\rangle\langle\psi;t\vert$, and 
the equation governing the dynamics of the Husimi function becomes
\begin{equation}
i\hbar\frac{\partial}{\partial t}Q_{\psi} = 
\text{tr}\left(\vert z\rangle\langle z\vert\widehat{H}\widehat{\rho} -
\widehat{\rho}\widehat{H}^{ \dagger}\vert z\rangle\langle z\vert\right).
\label{eq:time_derivative_Husimi_1}
\end{equation}

For hermitian Hamiltonians this is just von Neumann's relation cast in the
Husimi representation: $i\hbar\frac{\partial}{\partial t}Q_{\psi} = 
\text{tr}\left(\vert z\rangle\langle z\vert\left[\widehat{H},\widehat{\rho}\right]\right)$.
To further simplify this expression we assume that the Hamiltonian can be
expressed as a normal ordered power series on the creation and annihilation
operators as
\begin{equation}
\widehat{H}=\sum_{m,n}h_{mn}\widehat{a}^{\dagger m}\widehat{a}^{n}.
\label{eq:hamiltonian_taylor_series}
\end{equation}
If $h_{mn}=h_{nm}^{\star}$ the Hamiltonian is hermitian. The normalized matrix
elements of the Hamiltonian in the coherent states representation become 
\begin{equation}
H\left(z^{\prime\star},z\right)=
\frac{\langle z^{\prime}\vert\widehat{H}\vert z\rangle}{\langle z^{\prime}\vert z\rangle}
=\sum_{m,n}h_{mn}z^{\prime\star m}z^{n}.
\label{eq:hamiltonian_taylor_series_z}
\end{equation}

The action of the operators $\widehat{a}^{\dagger}$ and $\widehat{a}$ on the
projector $\vert z\rangle\langle z\vert$ is given by its differential algebra on
the representation of the coherent states \cite{gilmore75,zhang90,trimborn08}
and can be derived by appling them to the states as defined in equation
(\ref{eq:coherent_state_vacuum}). For the action to the right the 
following relations hold:
\begin{eqnarray*}
\widehat{a}\vert z\rangle\langle z\vert & = & z\vert z\rangle\langle z\vert,\\
\widehat{a}^{\dagger}\vert z\rangle\langle z\vert & = & \left(\frac{\partial}{\partial z}
+z^{\star}\right)\vert z\rangle\langle z\vert.
\end{eqnarray*}
The action of a general term of the Hamiltonian is given as
\begin{equation}
\widehat{a}^{\dagger i}\widehat{a}^{j}\vert z\rangle\langle z\vert=z^{j}
\left(\frac{\partial}{\partial z}+z^{\star}\right)^{i}\vert z\rangle\langle z\vert.
\label{eq:operator_right_action}
\end{equation}
The action to the left is given by analogous relations,
\begin{eqnarray*}
\vert z\rangle\langle z\vert\widehat{a} & = & \vert z\rangle\langle z\vert
\left(\frac{\overleftarrow{\partial}}{\partial z^{\star}}+z\right),\\
\vert z\rangle\langle z\vert\widehat{a}^{\dagger} & = & \vert z\rangle\langle z\vert z^{\star},
\end{eqnarray*}
leading to
\begin{equation}
\vert z\rangle\langle z\vert\widehat{a}^{\dagger j}\widehat{a}^{i}
=\vert z\rangle\langle z\vert
\left(\frac{\overleftarrow{\partial}}{\partial z^{\star}}+z\right)^{i}z^{\star j},
\label{eq:operator_left_action}
\end{equation}
which is just the hermitian transpose of (\ref{eq:operator_right_action})
in matrix notation. Eqs. (\ref{eq:operator_right_action})
and (\ref{eq:operator_left_action}) can be used in
Eq.(\ref{eq:time_derivative_Husimi_1}) describing the dynamics of the Husimi
function. Evaluating the first term inside the trace, using the Hamiltonian
power series (\ref{eq:hamiltonian_taylor_series}) leads to
\begin{eqnarray*}
\text{tr}\left(\vert z\rangle\langle z\vert\widehat{H}\widehat{\rho}\right) & 
= & \text{tr}\left(\vert z\rangle\langle z\vert\sum_{m,n}h_{mn}
\widehat{a}^{\dagger m}\widehat{a}^{n}\widehat{\rho}\right)\\
 & = & \text{tr}\left(\vert z\rangle\langle z\vert\widehat{\rho}\right)\sum_{m,n}h_{mn}
 \left[\left(\frac{\overleftarrow{\partial}}{\partial z^{\star}}+z\right)^{n}z^{\star m}\right]\\
 & = & \sum_{m,n}h_{mn}z^{\star m}\left(\frac{\partial}{\partial z^{\star}}+z\right)^{n}Q_{\psi}.
\end{eqnarray*}
The passing from the first to the second line can be accomplished
by the linearity of the trace, and from the second to the third line
a rearrangement of the factors was performed, putting the derivative operator
acting to the right as usual. A similar calculation can be done for
the second term inside the trace in Eq.(\ref{eq:time_derivative_Husimi_1}).
The resulting equation is 
\begin{equation}
i\hbar\frac{\partial}{\partial t}Q_{\psi} = \sum_{m,n}h_{mn}z^{\star
m}\left(\frac{\partial}{\partial z^{\star}}+z\right)^{n}Q_{\psi} -
\sum_{m,n}h_{mn}^{\star}z^{m} \left(\frac{\partial}{\partial z} +
z^{\star}\right)^{n}Q_{\psi}.
\label{eq:time_derivative_Husimi_2}
\end{equation}

Since the Husimi is a real function, $\frac{\partial}{\partial z^{\star}}
Q_{\psi} = \left(\frac{\partial}{\partial z}Q_{\psi}\right)^{\star}$. 
Therefore, although the dynamical equation for $Q_{\psi}$ is the sum of two
complex functions, its time evolution remains real as it should.

\subsection{Continuity equation and Flow}

In order to write the equation (\ref{eq:time_derivative_Husimi_2})
as a continuity equation all the derivatives with respect to $z^{\star}$
and $z$ must be put to the left, so that we can single out terms
of the form $\frac{\partial}{\partial z}J_{z}$ and $\frac{\partial}{\partial z^{\star}}J_{z^{\star}}$,
identifying in this way the Husimi currents $J_{z}$ and $J_{z^{\star}}$.
The relation between these complex currents in $\Xi_{z}$ and the real
ones in the classical phase space $\Xi_{cl}$ (\ref{eq:classical_currents_1})
can be obtained employing the following transformation law for the derivatives
in $\Xi_{z}$ and $\Xi_{cl}$:
\begin{equation}
\begin{gathered}\frac{\partial}{\partial
z}=\frac{\sqrt{\hbar/2}}{\gamma}\frac{\partial}{\partial
x}-i\gamma\sqrt{\hbar/2}\frac{\partial}{\partial p},\\
\frac{\partial}{\partial z^{\star}}=\frac{\sqrt{\hbar/2}}{\gamma}
\frac{\partial}{\partial x}+i\gamma\sqrt{\hbar/2}\frac{\partial}{\partial p}.
\end{gathered}
\label{eq:z_position_momentum_derivatives}
\end{equation}
We obtain
\begin{equation}
J_{x}=\frac{\sqrt{\hbar/2}}{\gamma}\left(J_{z^{\star}}+J_{z}\right),\qquad 
J_{p} = i\gamma\sqrt{\hbar/2}\left(J_{z^{\star}}-J_{z}\right).
\label{eq:currents_real}
\end{equation}

To put derivatives to the left in (\ref{eq:time_derivative_Husimi_2}), we need
to change the position of the terms $\left(\frac{\partial}{\partial
z^{\star}}+z\right)^{j}$ and $\left(\frac{\partial}{\partial
z}+z^{\star}\right)^{j}$ with that of the terms $z^{\star i}$ and
$z^{i}$, respectively. A concise way to express both changes is to define
\[
\begin{cases}
X=z^{\star} & \text{and }D=\left(\frac{\partial}{\partial z^{\star}}+z\right)\\
 & \text{or}\\
X=z & \text{and }D=\left(\frac{\partial}{\partial z}+z^{\star}\right).
\end{cases}
\]
In both cases, $XD=DX-1$, and we need to express the $X^{i}D^{j}$
term as a combination of those with the opposite ordering. This calculation
can be done by induction and the swapping of factors results
\[
X^{i}D^{j}=\sum_{k=0}^{\min\left(i,j\right)}r_{i,j,k}D^{j-k}X^{i-k},
\]
where
\[
r_{i,j,k}=\frac{\left(-1\right)^{k}i!j!}{k!\left(i-k\right)!\left(j-k\right)!}.
\]
Substituting the series above for the commutation of derivatives and
functions in Eq.(\ref{eq:time_derivative_Husimi_2}) and expanding the binomials
inside the definition of $D$, we end up with
\begin{multline}
\frac{\partial}{\partial t}Q_{\psi} =\frac{1}{i\hbar}\sum_{m,n}
\sum_{k=0}^{\min\left(m,n\right)}\sum_{l=0}^{n-k}
h_{mn}r_{m,n,k}\binom{n-k}{l}\frac{\partial^{l}}{\partial z^{\star l}}
z^{n-k-l}z^{\star m-k}Q_{\psi}+\\
-\frac{1}{i\hbar}\sum_{m,n}\sum_{k=0}^{\min\left(m,n\right)}\sum_{l=0}^{n-k}
h_{mn}^{\star}r_{m,n,k}\binom{n-k}{l}\frac{\partial^{l}}{\partial z^{l}}z^{\star n-k-l}z^{m-k}Q_{\psi}.
\label{eq:time_derivative_Husimi_3}
\end{multline}
Before we factor out the derivatives identifying the currents we notice that in
each summation there is a collection of terms having no derivatives at all, that
is, $l=0$. These terms play the role of a source contribution $\sigma$ to the
continuity equation. The explicit expression for this source is
\[
\sigma=\frac{1}{i\hbar}\sum_{m,n}\sum_{k=0}^{\min\left(m,n\right)}r_{m,n,k}
\left(h_{mn}z^{n-k}z^{\star m-k}-\text{c.c.}\right)Q_{\psi},
\]
where $\text{c.c.}$ stands for complex conjugate. If the Hamiltonian
is hermitian it can be shown, using the fact that the coefficients
$r_{i,j,k}$ are symmetric in the $i$ and $j$ indexes, that the
source vanishes, which is expected for the unitary evolution. 

From now on we assume that the Hamiltonian \emph{is} hermitian and
drop off the source terms. Taking out to the left one derivative of the
expression (\ref{eq:time_derivative_Husimi_3}) we can write at last
\begin{equation}
\frac{\partial}{\partial t}Q_{\psi}=-\frac{\partial}{\partial z}J_{z}-
\frac{\partial}{\partial z^{\star}}J_{z^{\star}},
\label{contf}
\end{equation}
where the currents are given by
\begin{eqnarray}
J_{z} & = & \frac{1}{i\hbar}\sum_{m,n}\sum_{k=0}^{\min\left(m,n\right)}\sum_{l=1}^{n-k}
h_{nm}r_{m,n,k}\binom{n-k}{l}\frac{\partial^{l-1}}{\partial z^{l-1}}z^{\star n-k-l}z^{m-k}Q_{\psi},
\label{eq:flow_definition_a}\\
J_{z^{\star}} & = & -\frac{1}{i\hbar}\sum_{m,n}\sum_{k=0}^{\min\left(m,n\right)}\sum_{l=1}^{n-k}
h_{mn}r_{m,n,k}\binom{n-k}{l}\frac{\partial^{l-1}}{\partial z^{\star l-1}}z^{n-k-l}z^{\star m-k}Q_{\psi}.
\label{eq:flow_definition_b}
\end{eqnarray}
It can be readily checked that $J_{z^{\star}}=J_{z}^{\star}$, such
that our expectation about having real currents (\ref{eq:currents_real})
on $\Xi_{cl}$ are met and the following relations stand
\[
J_{x}=\frac{\sqrt{2\hbar}}{\gamma}\text{Re}\left(J_{z}\right),\qquad J_{p}
=\gamma\sqrt{2\hbar}\text{Im}\left(J_{z}\right).
\]

As remarked before, these currents \emph{do} define a flow on the phase space,
and this flow has a crucial dependence on the shape of the Husimi function,
rendered by the high order derivatives appearing in
Eqs.(\ref{eq:flow_definition_a}) and (\ref{eq:flow_definition_b}),
coupled to the zero point Taylor coefficients of the Hamiltonian itself.
This dependence shows that the coupling in phase space does not have a local
character, thus being a fingerprint of nonlocality in this construction.

\section{Classical limit and Semiclassical corrections}

\subsection{Classical currents}

In the limit $\hbar\rightarrow 0$ the currents given by Eqs.
(\ref{eq:flow_definition_a}) and (\ref{eq:flow_definition_b}) should reduce to
the classical Liouville currents Eq.(\ref{eq:classical_currents_1}). A
complication that arises in the investigation of this limit is that the
Hamiltonian function $H\left(z^{\prime\star},z\right)$ itself involves terms of
order $\hbar$ or higher coming from the normal ordering process, and nothing
precludes the Husimi function $Q_{\psi}\left(z^{\star},z\right)$ of such $\hbar$
dependence as well. A true expansion of the equations in powers of $\hbar$ might
take these terms into account order by order. To avoid this extra difficulty we
will always take the full Hamiltonian $H\left(z^{\prime\star},z\right)$ and the
full Husimi function $Q_{\psi}\left(z^{\star},z\right)$ into account and expand
only the dynamical equation for the flow. Thus, each order will contain some
higher order terms in $\hbar$ coming only from the Hamiltonian and the Husimi
function. As an example consider
\begin{eqnarray}
\widehat{H}  =  \frac{\widehat{p}^{2}}{2m} -
\frac{k}{2}\widehat{x}^{2} + \lambda \widehat{x}^{4} + V_{0}
\label{hamil}
\end{eqnarray}
for which we find
\begin{eqnarray}
H  & =& - \frac{\hbar \Omega}{4}(z-z^\star)^2 - \frac{k \hbar}{4m\Omega}(z+z^\star)^2 +
\frac{\lambda \hbar^2}{4m^2\Omega^2}(z+z^\star)^4 + V_{0}\\
& & + \frac{\hbar \Omega}{4} - \frac{k \hbar}{4m\Omega} + \frac{3 \hbar^2 \lambda}{4m^2\Omega^2}
+ \frac{3 \hbar^2 \lambda}{2m^2\Omega^2}(z+z^\star)^2.
\end{eqnarray}

In equation (\ref{eq:z_position_momentum}) we factored out
the $\hbar$ dependence of the phase space variables $z$ and $z^{\star}$,
which makes them proportional to $\hbar^{-1/2}$. The
derivatives $\partial/\partial z$ and $\partial/\partial z^{\star}$, in turn,
are proportional to $\hbar^{1/2}$. In terms of $x$ and $p$,
we obtain
\begin{eqnarray}
\tilde{H}  & =& \frac{p^2}{2m} - \frac{k x^2}{2} + \lambda x^4 + V_{0} \\
& & + \frac{\hbar \Omega}{4} - \frac{\hbar k}{4m\Omega} + \frac{3 \hbar \lambda x^2}{m \Omega} +
\frac{3 \hbar^2 \lambda}{4m^2\Omega^2}
\end{eqnarray}
where the tilde identifies the functions written in the $\Xi_{cl}$ variables $x$
and $p$. 

In both these expressions, the first line corresponds to the {\it classical}
Hamiltonian and is $\hbar$ independent. The second lines contain the
corrections. It is therefore clear that if the quantum Hamiltonian operator is
independent of $\hbar$, each monomial  $h_{mn}z^{\star m}z^{n}$ is of order
$\hbar^0$ plus corrections coming from normal ordering the creation and
annihilation operators.  

In what follows we investigate the semiclassical limit without expanding the
Hamiltonian and the Husimi function. Although such complete expansion could be
performed, it is much more complicated and does not bring any insight into the
structure of the flow. Using this scheme, the classical limit becomes
\[
\lim_{\hbar\rightarrow0}J_{z}=\lim_{\hbar\rightarrow0}\sum_{q=0}^{q_{max}}
\hbar^{q}j_{q}\left(z^{\star},z;t\right)
=j_{0}\left(z^{\star},z;t\right),
\]
with all terms $h_{mn}z^{\star m}z^{n}$ treated as $\hbar^0$. Here 
$q_{max}$ as the highest power of $\hbar$ in such expansion. In order to 
connect $q$ with the indexes $m$, $n$, $k$ and $l$ present in the summation 
formulae for the currents we write Eq.(\ref{eq:flow_definition_a}) as
\[
J_{z} = i\sum_{m,n} \sum_{k=0}^{\min\left(m,n\right)}
\sum_{l=1}^{n-k}\hbar^{k+l-1}f_{m,n, k,l},
\]
set $q=k+l-1$ and identify $j_{q}$ with the proper summation over a set of 
the $f_{m,n,k,l}$ (see below). For hermitian Hamiltonians whose highest power in
the annihilation and creation operators is $n_{max}$, we find
$q_{max}=n_{max}-1$

To analyze the general term with a given $\hbar$ power $q$ in
Eq.(\ref{eq:flow_definition_a}), we replace the $l$ by $l=q -k+ 1$, with $k\leq
q\leq n-1$, whenever $k\leq n-1$:
\begin{equation}
J_{z}=\frac{1}{i\hbar}\sum_{m,n}\sum_{k=0}^{\min\left(m,n-1\right)}
\sum_{q=k}^{n-1}h_{nm}r_{m,n,k}\binom{n-k}{q-k+1}
\frac{\partial^{q-k}}{\partial z^{q-k}}z^{\star n-q-1}z^{m-k}Q_{\psi}.
\label{eq:hbar_expansion_general_term}
\end{equation}
A similar expression holds for $J_{z^{\star}}$. Now the $\hbar\rightarrow0$
limit can be taken by selecting the term $q=k=0$ in this summation. The
result is
\begin{eqnarray}
J_{z}  =  \frac{1}{i\hbar}\sum_{m,n}nh_{nm}z^{\star n-1}z^{m}Q_{\psi} = 
\frac{1}{i\hbar}\frac{\partial H}{\partial z^{\star}}Q_{\psi},
\label{eq:classical_current_z_a}
\end{eqnarray}
where $H\equiv H\left(z^{\star},z\right)$, as defined in
(\ref{eq:hamiltonian_taylor_series_z}), does contain higher powers of $\hbar$ as
discussed above, as the Husimi function. Analogously, 
\begin{equation}
J_{z^{\star}}=-\frac{1}{i\hbar}\frac{\partial H}{\partial z}Q_{\psi}.
\label{eq:classical_current_z_b}
\end{equation}

The classical currents, calculated from (\ref{eq:classical_current_z_a}) and
(\ref{eq:classical_current_z_b}) are given by
\begin{equation}
J_{x}=\frac{\partial\tilde{H}}{\partial p}\tilde{Q}_{\psi},\qquad 
J_{p}=-\frac{\partial\tilde{H}}{\partial x}\tilde{Q}_{\psi}.
\label{eq:classical_currents_2}
\end{equation}
By direct comparison of equations (\ref{eq:classical_currents_1})
and (\ref{eq:classical_currents_2}) we can extract the classical
equations of motion in the phase space
\[
\dot{x}=\frac{\partial\tilde{H}}{\partial p},\qquad
\dot{p}=-\frac{\partial\tilde{H}}{\partial x}.
\]
Notice that we have \emph{not} made any assumptions regarding trajectories
being guided by a Hamiltonian. The classical structure emerges naturally from
the quantum case. Going a step further, the dynamics of the Husimi function in
the classical limit is governed by the equation
\begin{eqnarray*}
\frac{\partial}{\partial t}\tilde{Q}_{\psi} & = & -\frac{\partial}{\partial x}
\left(\frac{\partial\tilde{H}}{\partial p}\tilde{Q}_{\psi}\right)-\frac{\partial}{\partial p}
\left(-\frac{\partial\tilde{H}}{\partial x}\tilde{Q}_{\psi}\right)\\
 & = & \left\{ \tilde{H},\tilde{Q}_{\psi}\right\} _{\left[x,p\right]},
\end{eqnarray*}
where $\left\{ \cdot,\cdot\right\} _{\left[x,p\right]}$ is the Poisson
bracket in the coordinates $x$ and $p$. The additional terms, in higher
order of $\hbar$, in Eq.(\ref{eq:hbar_expansion_general_term})
lead to quantum deviations from the classical flow.

\subsection{Semiclassical corrections}

The first quantum corrections to the classical dynamics are given by the $q=1$
terms in Eq.(\ref{eq:hbar_expansion_general_term}). In this case, two
contributions arise, from $k=0$ and $k=1$. For $J_{z}$ and $J_{z^{\star}}$ this
amounts to
\begin{eqnarray*}
J_{z}|_{q\leq1} & = & \frac{1}{i\hbar}\frac{\partial H}{\partial z^{\star}}Q_{\psi} + 
\frac{1}{i\hbar}\frac{1}{2}\frac{\partial^{2}H}{\partial z^{\star2}} 
\frac{\partial}{\partial z}Q_{\psi}-\frac{1}{i\hbar}\frac{1}{2}
\frac{\partial^{3}H}{\partial z^{\star2}\partial z}Q_{\psi},\\
J_{z^{\star}}|_{q\leq1} & = & -\frac{1}{i\hbar}\frac{\partial H}{\partial z}Q_{\psi} - 
\frac{1}{i\hbar}\frac{1}{2}\frac{\partial^{2}H}{\partial z^{2}}
\frac{\partial}{\partial z^{\star}}Q_{\psi} +
\frac{1}{i\hbar}\frac{1}{2}\frac{\partial^{3}H}{\partial z^{2}\partial
z^{\star}}Q_{\psi}.
\end{eqnarray*}
This correction lends the flow exact for Hamiltonians quadratic in the operators
$\widehat{a}^{\dagger}$ and $\widehat{a}$, $n_{max}=2$. Thus $q_{max}=n_{max}
-1 = 1$ is the highest correction for quadratic Hamiltonians, which, therefore, 
may be termed the semiclassical approximation for the current. Substituting
these expressions into Eq.(\ref{contf}) we find
\begin{eqnarray*}
\frac{\partial}{\partial t}Q_{\psi}|_{q\leq1} & = & \frac{1}{i\hbar}\left\{
H,Q_{\psi}\right\} _{\left[z,z^{\star}\right]} +
 \frac{1}{i\hbar}\frac{1}{2}\frac{\partial^{2}H}{\partial z^{2}}
\frac{\partial^{2}}{\partial z^{\star2}}Q_{\psi} - 
\frac{1}{i\hbar}\frac{1}{2}\frac{\partial^{2}H}{\partial z^{\star2}}
\frac{\partial^{2}}{\partial z^{2}}Q_{\psi}.
\end{eqnarray*}
This is an anisotropic diffusion equation, which can be related to the thawing
of a wavepacket over the phase space. Two simple examples, for the sake of
illustration, are:

i) The harmonic oscillator of mass $m$ and frequency $\Omega$. The
Hamiltonian is
\[
\widehat{H}_{HO}=\hbar\Omega\left(\widehat{a}^{\dagger}\widehat{a}+\frac{1}{2}\right),
\]
and the diffusive correction is zero. This means that the Husimi
function for this system does not spread, following the classical
flow.

ii) The free particle with mass $m$,
\[
\widehat{H}_{FP}=-\frac{\hbar\Omega}{4}\left(\widehat{a}^{\dagger2} +
\widehat{a}^{2}-2\widehat{a}^{\dagger}\widehat{a}-1\right),
\]
and the diffusive correction leads to
\[
\frac{\partial}{\partial t}\tilde{Q}_{\psi}|_{q\leq1}
=-\frac{p}{m}\frac{\partial}{\partial x}\tilde{Q}_{\psi}+\frac{\hbar\Omega}{2}
\frac{\partial^{2}}{\partial x\partial p}\tilde{Q}_{\psi},
\]
where it is possible to identify the classical velocity $p/m$ and
the diffusion coefficient $\hbar\Omega$, which depends on parameters
of the states used in the representation.

\section{Husimi Flow for a Particle in a Double Well}

In this section we illustrate the main features of the Husimi flow with a
numerical example. In particular, we are going to evaluate higher order
semiclassical terms in the dynamical equation for the flow,
Eq.(\ref{eq:hbar_expansion_general_term}). 

\begin{figure}[!ht]
\begin{centering}
\includegraphics[scale=0.4]{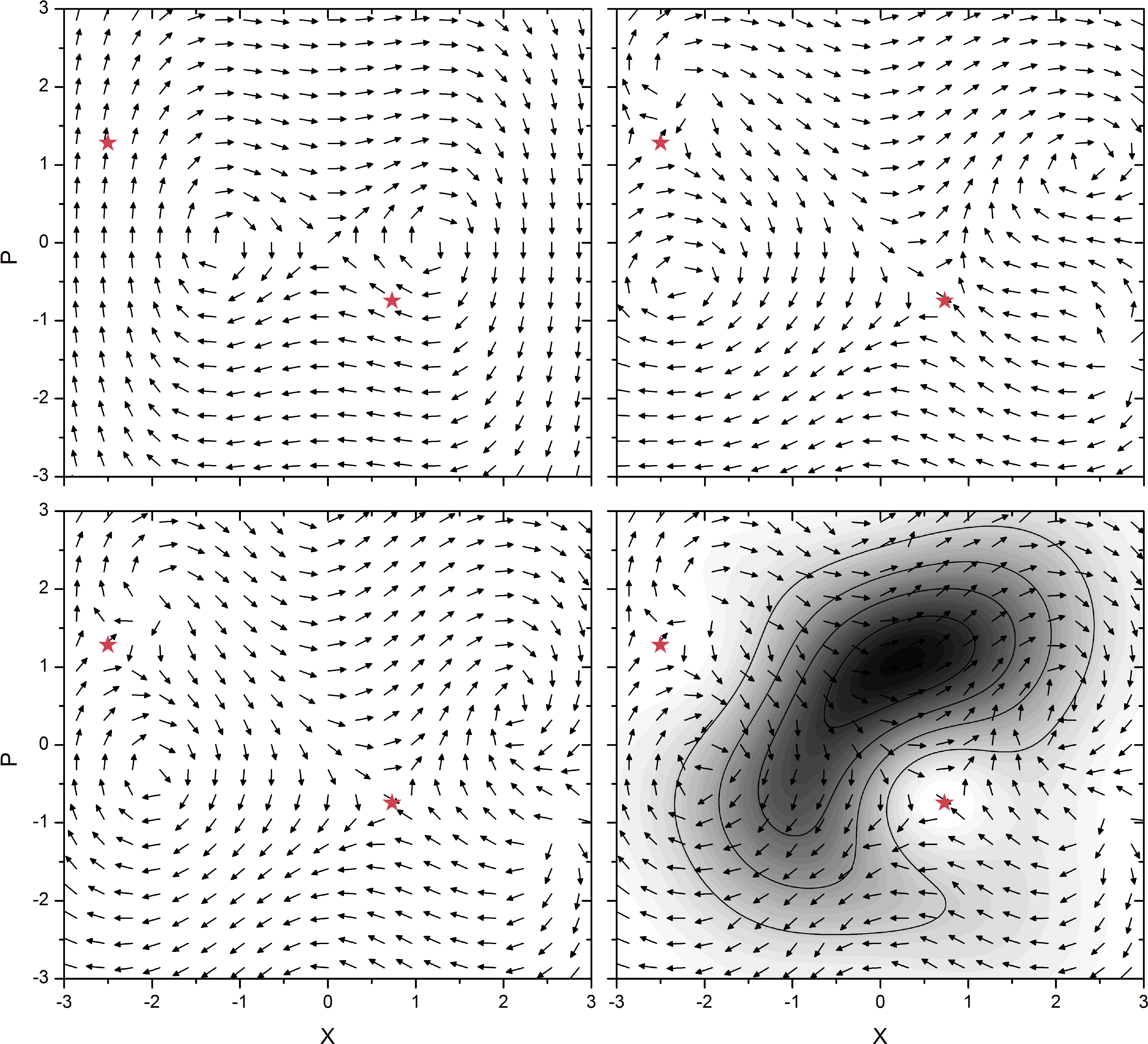}
\par\end{centering}
\caption{(Color online) Husimi flow at $t=3.5$. The unit arrows indicate the
direction of the current, regardless of the intensity. Upper left: $\hbar^{0}$
(classical flow); upper right: $\hbar^{1}$ correction; lower left: $\hbar^{2}$
correction; lower right: $\hbar^{3}$ correction (exact Husimi flow). The red
stars mark the zeros of the Husimi function (plotted as background in the last
image (white for 0 and dark gray for highest intensity).}
\label{fig1}
\end{figure}

Although the currents are defined by an \emph{a priori} infinite series of
terms, comprising the Taylor expansion of the Hamiltonian, the series truncate
for polynomial potentials. In order to obtain exact results we choose as
toy-model the Hamiltonian (\ref{hamil}) describing a particle in a symmetric
double well potential:
\begin{displaymath}
\widehat{H}  =  \frac{\widehat{p}^{2}}{2m}+
\lambda \widehat{x}^{4}-\frac{k}{2}\widehat{x}^{2}+V_{0}.
\end{displaymath}
The Hamiltonian function can be rearranged as 
\begin{eqnarray}
H\left(z^{\star},z\right) =  -\frac{\hbar\Omega}{2}\left(z^{\star 2}+z^{2}\right) 
+\frac{\lambda\hbar^{2}}{4m^{2}\Omega^{2}}\left(z^{\star}
+z\right)^{4} + 2V_{0},
\label{eq:toy_model_hamiltonian}
\end{eqnarray}
where the parameters of the potential are set to be $k=\left(m\Omega^{2} + 
\frac{6\lambda\hbar}{m\Omega}\right)$ and 
$V_{0}=\frac{3\lambda\hbar^{2}}{4m^{2}\Omega^{2}}$. The classical
currents for the averaged Hamiltonian $H\left(z^{\star},z\right)$
have 3 stationary points, all of them with $\text{Im}\left(z\right)=0$:
one saddle at $\text{Re}\left(z\right)=0$ and two clockwise centers
at $\text{Re}\left(z\right)=\pm\sqrt{\frac{m^{2}\Omega^{3}}{8\lambda\hbar}}$ as
shown in Fig.\ref{fig1} (upper left corner).

For the numerical calculations we used $2m=3\lambda=\Omega/2=\hbar=1$, and the
initial wavepacket is a coherent state $\vert z_{0}\rangle$ with 
$z_{0}=0.9228 i-0.866$, corresponding to the center
of the left well with energy equal to about two times the classical energy to
pass over the central barrier. The time evolution of the wave
packet was performed with the Split-Time-Operator method using Fast Fourier
Transforms between position and momentum representations \cite{bandrauk93}. 

Figure \ref{fig1} shows the direction of the flow at time $t=3.5$ using
increasingly accurate approximations. For the hamiltonian
(\ref{eq:toy_model_hamiltonian}) the highest $\hbar$ power in the flow series is
$q_{max}=3$ and the panels show the classical flow $q=0$, the first order
semiclassical correction $q=1$, the next order $q=2$ and the exact result
corresponding to $q=3$, for which the Husimi function is also shown as a grey
scale contour plot. The corrections in increasing powers of $\hbar$ change the
overall structure of the current portraits. The center stationary
points present in the classical flow are displaced from their locations both in
momentum and position. It can also be seen that additional critical points
appear close to the zeros of the Husimi function, and exactly at the zeros
they are saddle points of the flow.

\begin{figure}[!ht]
\begin{centering}
\includegraphics[scale=0.4]{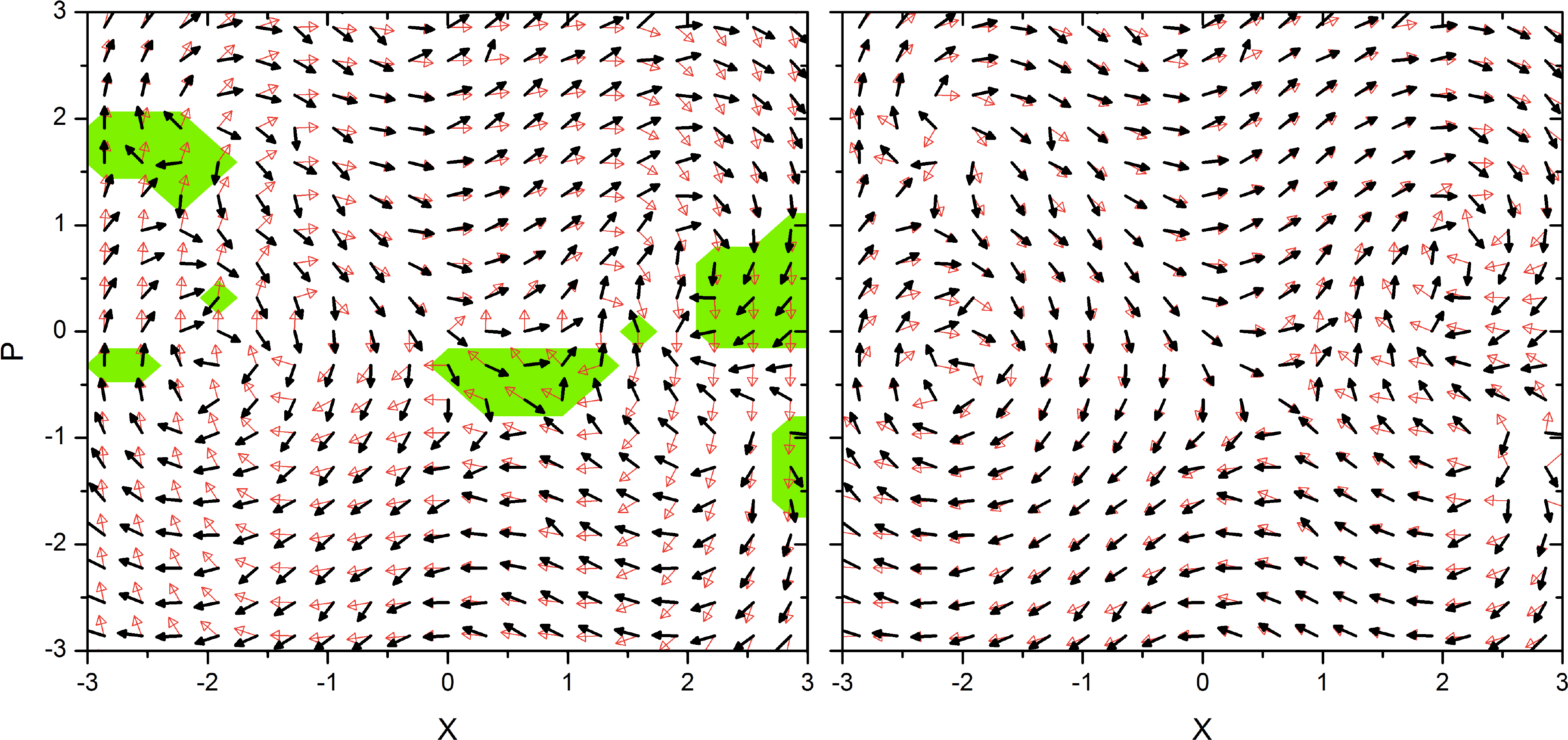}
\par\end{centering}
\caption{(Color online) Comparison of flow vectors. Thick black arrows are the exact
currents and thin red arrows show lower order corrections: $q=0$ on the left and
$q=1$ on the right. The shaded green areas in the left panel show regions where
momentum inversion occurs.}
\label{fig2}
\end{figure}

Figure \ref{fig2} shows a comparison between the exact and classical flows (left
panel) and between the exact and the first order correction (right panel).
Except for some small displacements, the $\hbar^{1}$ correction almost mimics
the exact quantum flow.

Classically, a particle moves towards the positive position direction if it has
a positive momentum and vice-versa. This can be clearly seen in
Fig.\ref{fig1}(a): if $p>0$  the flow is to the right and if  $p<0$ it points to
the left. The quantum flow, on the other hand, Fig. \ref{fig1}(d) does not
follow that rule everywhere in phase space. There are regions (shaded green
areas in Figure \ref{fig2}) where $p>0$ ($p<0$) that have flow lines pointing 
towards the left (right), the classically wrong direction. This
counter-classical motion can be interpreted as an evidence of tunneling
represented in this phase space formulation. 

\begin{figure}[!ht]
\begin{centering}
\includegraphics[scale=0.4]{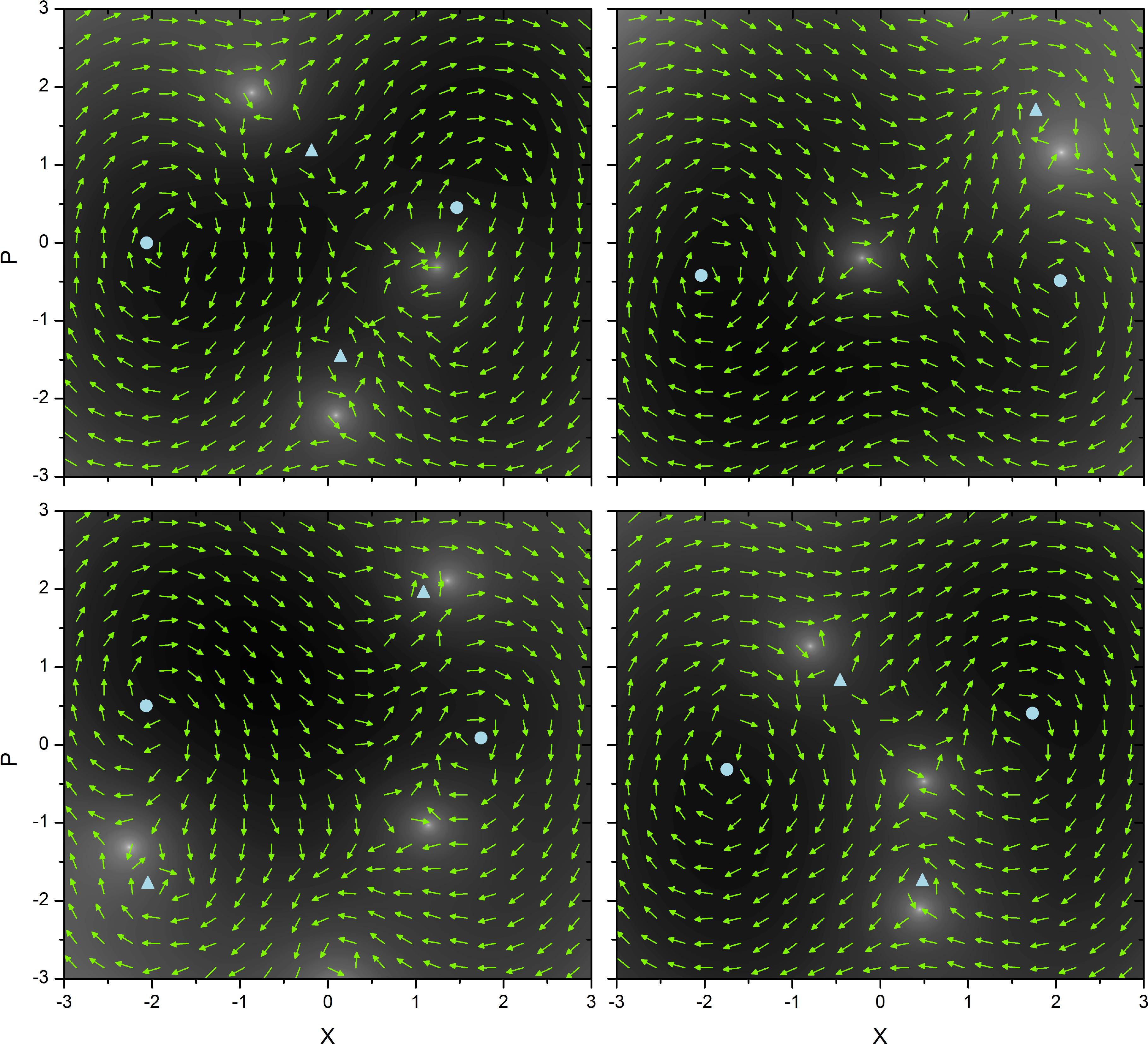}
\par\end{centering}
\caption{(Color online) Logarithmic plot of the Husimi function superposed with
the current unit vectors. White denotes small values ($10^{-10}$) and black the
highest ($10^{0}$) for the function. Upper left $t=1.0$, upper right $t=2.0$,
lower left $t=3.0$ and lower right $t=4.0$. The zeros are shown by the
white spots. Except for one zero, all the others are followed by a flow center
(blue triangles). The shifted classical centers are also indicated (blue
circles).}
\label{fig3}
\end{figure}

Figure \ref{fig3} shows the zeros and the flow of the Husimi function for the  
double well potential  at different times. The zeros of the Husimi function
completely characterize a quantum state
\cite{leboeuf90,cibils92,nonnen97,korsch97}. For eigenstates the zeros are
static (the so called stellar representation), but for general states the zeros
usually display a non-trivial dynamics. Absent in the initial state, the zeros
approach the region near the wells coming from infinity and move along
trajectories that are neither classical nor follows the Husimi flow. We
identified that the zeros of the Husimi function correspond to saddle points 
of the Husimi flow, as seen in Figure \ref{fig3}, and found no other saddle
points except those located over the Husimi zeros. We speculate that for every
additional zero of the Husimi function there must exist a center of stability in
order to conserve the total index of the flow on the phase space (Figure
\ref{fig3}).  

\section{Final remarks}

In this work we constructed the phase space flow for the Husimi representation
of a one-particle quantum state. The formulas were tested for a particle in a
double well potential. We showed that the flow lines differ substantially
from its corresponding classical structure, displacing the stagnation points and
modifying their quantity. We also observed momentum reversion of the currents in
some areas of the phase space. These reversions had already been observed for
the Wigner flow, although there is a difference regarding their interpretation
in each of these formulations. The Wigner reversion was conjectured to be caused
by the function's non-posivity, which is usually related to quantumness
\cite{skr13}. As the Husimi function is positive definite this explanation does
not work here and we claim that momentum reversion is a more robust blueprint of
quantumness than the negativity of the representation itself (see also
\cite{ferrie11}), since it is observed in  both representations. 

Contrary to the Wigner flow \cite{skr13}, the Husimi flow does not allow for the
birth or merging of critical points. This seems to be a consequence of the
isolation of the Husimi zeros, since they appear to be always associated to
saddle points of the flow. For small propagation times the zeros move from
infinity into the region where the Husimi function is significant and remain
there without bifurcating. Also, the Husimi zeros are followed by a center
stagnation point, which appears due to index conservation of the flow.

\section*{Acknowledgements}

It is a pleasure to thank A. Grigolo and T. F. Viscondi for helpful comments. 
This work was partly supported by FAPESP and CNPq.

\end{document}